\def\breakon{\end{multicols}\widetext\vspace{-.2cm} 
\noindent\rule{.48\linewidth}{.3mm}\rule{.3mm}{.3cm}\vspace{.0cm}} 
\def\breakoff{\vspace{-.2cm} 
\noindent 
\rule{.52\linewidth}{.0mm}\rule[-.27cm]{.3mm}{.3cm}\rule{.48\linewidth}{.3mm} 
\vspace{-.3cm} 
\begin{multicols}{2} 
\narrowtext} 
\def\beq{\begin{equation}} 
\def\enq{\end{equation}} 
\def\bqa{\begin{eqnarray}} 
\def\eqa{\end{eqnarray}} 

\documentstyle[aps,prb,multicol,epsfig]{revtex} 

\begin{document} 
 
\draft 

\widetext 
 
\title{Long-Range Charge Transfer in DNA through Polaron Diffusion} 
 
\author{Chun-Min Chang$^1$, A. H. Castro Neto$^{2,*}$,  A. R. Bishop$^3$}

\address{$^1$ Department of Physics, University of California, Riverside, CA 92521 \\
$^2$ Department of Physics, Boston University, 590 Commonwealth Ave. \\
Boston, MA 02215 \\
$^3$ Theoretical Division and Center for Nonlinear Studies, \\
Los Alamos National Laboratory, Los Alamos, New Mexico 87545} 
 
\date{\today} 
 
\maketitle 
 
  
\begin{abstract} 
 Recent experimental evidence shows that the $\pi $ orbitals along the 
stacking of base pairs can facilitate the long-range charge transfer in DNA
[5-7]. Proton motion in the base pair hydrogen bonds has also been 
found to affect the transfer rate. To explain this behavior we propose a 
model considering interactions of doped charges with hydrogen bonds and 
vibrations in DNA. The charge trapped by either protons or vibrations 
can cause structural distortions leading to polaron formation. 
By further considering polaron diffusion in DNA we 
find 
that the charge transfer rate derived from the diffusion coefficient is in 
agreement with the experimental results [33].
\end{abstract} 
 
 
 
 
\begin{multicols}{2} 
 
\narrowtext 
 
\section{Introduction} 
\label{sec:Intro} 
 
The question on whether DNA is a conductor or an insulator is still
controversial due to the different results obtained in the measurements of DNA
conductivity \cite{Fink,Porath,Kasumov,Gruner}. The related issue of
long-range charge transfer in DNA which is associated with the problem of
charge mobility has also been discussed for many years with an equal amount of
heated debate. It has been suggested that the overlap of the electronic $\pi 
$ orbitals along the stacked base pairs provides a pathway for charge
propagation over 50 \AA\ and the reaction rate between electron donors and acceptors
does not decay exponentially with the distance \cite{Meggers,Nunez & Barton,Hall & Barton}.

A multiple-step hopping mechanism was recently proposed to explain the
charge transfer behavior in DNA \cite{Jortner}. In this theory, the single
G-C base pair is considered as a hole donor due to its low ionization
potential if compared to the one on A-T base pairs. Long-range charge transfer is
accomplished by a series of incoherent charge tunneling events between two
nearest G-C base pairs separated by A-T pairs. The probability of the
short-range tunneling (super-exchange) strongly depends on the
distance between the two G-C base pairs. However, recent experimental
results for the reaction rate only showed a very slight distance dependence
for the charge hole transfer through more than three A-T base pairs \cite
{Giese nature}. It is now believed that a hole can be created in a A-T base
pair by thermal activation from a G-C base pair when charge tunneling
becomes unlikely over a long distance \cite{Giese and
Spichey,Berlin,Bixon}. The long-range hopping can, therefore, be carried out
through
a long A-T bridge just as by a series of short-range tunnelings in the G-C
pairs. Although this mechanism seems to give a good interpretation of the
charge transfer in DNA, the cause of the incoherent charge hopping over
localized states of the hole carriers with such low reaction rate ($%
10^{9}\sim 10^{6} s^{-1}$) has not yet been well understood.

Besides the multiple-step hopping mechanism, polaron motion has also been
considered as a possible mechanism to explain the phenomena of DNA charge
transfer \cite{Schuster,Conwell}. The charge coupling with the DNA
structural deformations can create a polaron and cause a localized state.
From the study of the dynamical properties in one dimensional systems, it is
known that the polaron behaves as a Brownian particle that collides with the
low energy excitations of its environment which acts as a heat bath \cite
{Castro Neto}. This diffusive behavior is very similar to the multiple-step
hopping mechanism based on a random walk model, although the former occurs
in a continuous media while the latter is considered on the discrete sites
of a lattice. Thus, the behavior of the incoherent charge hopping can be
understood as polaron diffusion.

For most physical systems acoustical and optical phonons are the main
lattice excitations. However, when a system contains hydrogen bonds, proton
motions also need to be considered. Instead of oscillatory motions in a
single-minimum potential, protons can tunnel from one side of a hydrogen
bond to another in a double-minimum potential \cite{Kong}. This proton
tunneling causes interstrand charge hopping and is important in
spontaneous mutations. Since genetic information can be preserved only when
a nucleotide base is matched with its complementary one by hydrogen 
bonding, the
positions of protons are the main identification for DNA polymerases to make
replications with high fidelity. However, the reaction of proton transfer
could generate tautomeric base pairs and destroy the fidelity \cite{Watson}.
The mechanism of the proton tunneling in the isolated base pairs and its
possible biological implications were widely discussed by L\"{o}wdin many
years ago \cite{Lowdin}. It has also been suggested that proton transfer and
charge conduction in stacked base pairs are affected by each other \cite
{Colson,Faraggi,Steenken}. The recent experimental results on the influence
of mismatched base pairs show that the proton transfer is required for the
long-range charge transfer \cite{Giese}. Thus the effect of proton tunneling
will be discussed here.

The paper is organized as follows. The model for charge motion in DNA will
be introduced in the next Section, considering charge hopping,
possible fluctuations in structure and their interactions. We will present
the formation of a polaron in Section III, together with the study of
polaron stability and proton delocalization in the hydrogen bonds. In 
Section IV, we calculate the polaron diffusion coefficient and the results
will be compared with the experimental reaction rate of charge transfer in
DNA. Section V contains our conclusions.

\section{Theoretical Model} 
 
\begin{figure} 
\hspace{-.2cm} 
\epsfxsize=8cm 
\vspace{.5cm} 
\epsfbox{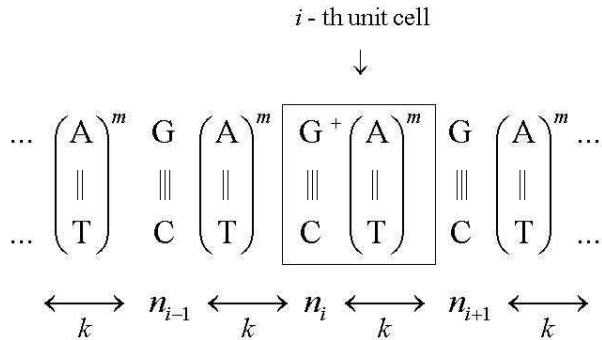} 
\caption{Charge hopping along the stacked base pairs of a periodic DNA
chain. $%
n_{i}$ is the number of the charges in the i-th unit cell.}
\label{pdfig1} 
\end{figure}

Although biologically functional DNA is an aperiodic system, we will focus
here 
on a periodic DNA structure. Since the polaron state discussed here is
highly localized (see below), we believed that the periodicity condition is
not very constraining when we consider charge transfer over intermediate
distance. In fact, in some experimental setups one looks for charge transfer
between two G-C pairs which are separated by many (periodic) A-T pairs. As
we can see in Fig. $\ref{pdfig1}$, following hole injection 
into the system, the
charge localizes in the G-C base pair but can migrate from one unit cell to
another. The essential ingredients in our model are ($i$) the electronic
system, that realizes the charge conduction;\ ($ii$) the possible structural
fluctuations, which include phonons and proton motion in hydrogen bonds; and
($iii$) the coupling between the charges and the structure.

For a non-interacting electron system, the tight-binding model is
extensively used in many theories of DNA charge transport \cite
{Berlin,Conwell,Grozema}: 
\begin{equation}
H_{e}=\sum_{n}-t_{0}(C_{i}^{\dagger }C_{i+1}+C_{i+1}^{\dagger }C_{i})
\label{tight-binding model}
\end{equation}
where $C_{i}$($C_{i}^{\dagger }$) is the electron annihilation(creation)
operator at the $i$-th site and $t_{0}$ is the overlap integral of
electronic $\pi $ orbitals between two nearest neighboring sites. (We
disregard the electron spin since we are not describing magnetic phenomena
here.)
The energy spectrum of the electrons is $\hbar \omega _{k}=-2t_{0}\cos
\left( ka\right) $ giving a bandwidth $4t_{0}$\ , where $a$ is the lattice
spacing and the wave vectors $k$ is $2\pi n/L$ with $n$ as an integer and the
system size $L$. In the continuum limit ($a\rightarrow 0$), the Hamiltonian
describes the kinetic energy of charge particles with a Bloch wavefunction
and effective mass $m$ , where $\hbar ^{2}/2ma^{2}$ is used to replace the
overlap integral $t_{0}$ of the discrete model.

For the phonon part of the structural fluctuations in DNA, we consider that
the relative motions between two different base pairs can be represented by
an acoustical phonon mode and the vibrational motion inside a base pair by
optical phonons \cite{Kittel}. These modes represent lattice distortions,
such as sliding, twisting or bending. The Hamiltonian that describes these
modes is:
\begin{eqnarray}
H_{ph}&=&\sum_{i}\left[ \frac{p_{i}^{2}}{2M}+\frac{1}{2}M\omega _{s}^{2}\left(
u_{i+1}-u_{i}\right) ^{2}\right] 
\nonumber
\\
&+&
\sum_{i}\left[ \frac{P_{i}^{2}}{2M}+\frac{1%
}{2}M\omega _{0}^{2}v_{i}^{2}\right]  \, ,
\label{phonons}
\end{eqnarray}
where $u_{i}$\ and $v_{i}$ are the lattice displacement and the internal
vibration coordinate of the $i$-th unit cell, $p_{i}$\ and $P_{i}$\ are
their conjugated momentum, respectively, $M$ is the mass of a unit cell, $%
\omega _{0}$ the oscillation frequency of the optical phonon and the
dispersion relation of the acoustical motion is: $\omega _{s}\left( k\right)
=c_{s}k$ , where $c_{s}$ is the sound velocity along the chain.

The charge coupling to acoustical phonons is given by the
Su-Schrieffer-Heeger model \cite{Su} and the interaction with optical
phonons is described via the molecular crystal model of Holstein \cite
{Holstein}. Thus the total electron-phonon interactions of DNA system is 
\begin{eqnarray}
H_{e-ph}&=&\sum_{i}\frac{\gamma _{s}}{a}\left( u_{i+1}-u_{i}\right)
(C_{i}^{\dagger }C_{i+1}+C_{i+1}^{\dagger }C_{i})
\nonumber
\\
&+&\sum_{i}\gamma
_{v}v_{i}C_{i}^{\dagger }C_{i}\text{ ,}  \label{phonon interactions}
\end{eqnarray}
where $\gamma _{s}$\ and $\gamma _{v}$ are the coupling constants. In
systems with half-filled conduction band, such as an one-dimensional $\pi $%
-conjugated polymers, the SSH term generates dimerization in the ground
state (Peierls instability) and forms 
solitons in the excited states \cite{Heeger}%
. However, for DNA, considered here as a band insulator, both
interactions can generate lattice distortions and lead to polaron formation
when a charge is doped into the molecule. The second term in (\ref{phonon
interactions}) usually induces small polarons in ionic crystals.

\begin{figure} 
\hspace{-.2cm} 
\epsfxsize=8cm 
\vspace{.5cm} 
\epsfbox{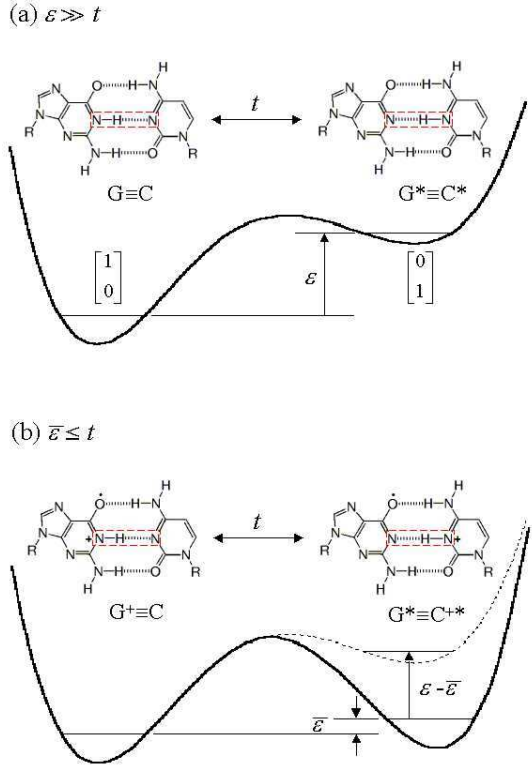} 
\caption{Proton transfer in the hydrogen bond of a DNA base pair described
by a two-level system. (a) For a neutral G-C base pair, proton transfer is
unlikely to cause the tautomeric form and we have $\protect\varepsilon \gg t$
. (b) The radical cation of a G-C base pair only has a little energy
difference with the tautomeric form, i.e., 
$\bar{\protect\varepsilon} < t$.}
\label{pdfig2} 
\end{figure}

For the proton motion in the hydrogen bonds we use a two-level system to
describe the tunneling behavior (see Fig. $\ref{pdfig2}$) \cite{Leggett}: 
\begin{equation}
H_{\sigma }=\frac{1}{2}\left( -\varepsilon \sigma ^{z}+t\sigma ^{x}\right) ,
\label{two-level system}
\end{equation}
where $\sigma ^{z}$ and $\sigma ^{x}$ are Pauli matrices, $\varepsilon $ is
the energy bias between the two localized proton states and $t$ is the
tunneling matrix element. The ratio $t/\varepsilon $ provides information of
the proton motion in the system. When the ratio is small, the protons are
localized in one side of the hydrogen bonds. However, when it is large, the
protons are delocalized. As we can see in Fig. $\ref{pdfig2}$(a), the normal
G-C base pair is the lower energy state and its tautomeric form (G*$\equiv $%
C*) is an excited state. When $\varepsilon \gg t$ , the probability of
having a tautomeric form is extremely small. However, in Fig. $\ref{pdfig2}$%
(b), the radical cation of a G-C base pair has almost the same energy as its
tautomeric form$\ $\cite{Hutter}, i.e., $\bar{\varepsilon}\lesssim t$\ . In
this case, the proton state becomes delocalized in the hydrogen bonds.

To model the coupling between the protons in the hydrogen bond and the
charges in the DNA strand, we use: 
\begin{equation}
H_{e-\sigma }=\gamma _{\sigma }\left( \sigma _{i}^{z}-1\right) n_{i}\text{ ,}
\label{proton interaction}
\end{equation}
where $\gamma _{\sigma }$ is the coupling intensity and $n_{i}=C_{i}^{%
\dagger }C_{i}$ is the number of charges on the $i$-th site. When the proton
is in the lower energy state ($\sigma _{i}^{z}=1$) or there is no charge
around the hydrogen bond ($n_{i}=0$), the coupling vanishes. Although there
are many hydrogen bonds in the base pairs of a unit cell, it is not
necessary to consider all of the possible proton transfers since at room
temperature the transitions to the tautomeric forms are very unlikely to
occur. We only need to consider the hydrogen bond with the most probable
proton transfer with charge injection into the unit cell, i.e., only one
two-level system is required for each unit cell.

By giving a total Hamiltonian as the sum of (\ref{tight-binding model}), (%
\ref{phonons}), (\ref{phonon interactions}), (\ref{two-level system}) and (%
\ref{proton interaction})\, we have our complete model for charge motion in
DNA.

\section{Formation of Polaron in DNA}

Although polaron formation has already been considered very important for
charge motion in DNA \cite{Schuster,Conwell,G Schuster}, the effect of the
proton motion in the hydrogen bonds has not yet been discussed in the
literature. Here, to investigate this factor, we analytically solve our
model by taking the Hamiltonians into the continuum limit. The carrier
wavefunction $\psi \left( x,t\right) $ and the phonon modes $u_{s}\left(
x,t\right) $,\ $u_{v}\left( x,t\right) $ are considered as functions of
time and a continuous coordinate $x$ .The equations of motion for the
variables in the problem can be studied in the semiclassical limit ($\hbar
\rightarrow 0$) and by a change of variables $\lambda =x-\upsilon t$ , where 
$\upsilon $ is the polaron velocity. In this case, we have \cite{Castro Neto}:
\begin{equation}
\begin{array}{l}
u_{s\left( v\right) }\left( x,t\right) =u_{s\left( v\right) }\left(
x-\upsilon t\right) =u_{s\left( v\right) }\left( \lambda \right) \\ 
\psi \left( x,t\right) =\phi _{0}\left( \lambda \right) \exp \left[ \frac{i}{%
\hbar }\left( m\upsilon x-E_{0}t\right) \right] ,
\end{array}
\label{Lorentz transformation}
\end{equation}
where $E_{0}$ is the binding energy of the polaron, and the problem reduces
to a nonlinear Schr\"odinger equation for the wavefunction $\phi _{0}$\ : 
\begin{eqnarray}
\frac{\hbar ^{2}}{2m}\frac{d^{2}\phi _{0}}{d\lambda ^{2}}&+&\left( E_{0}-
\frac{
m\upsilon ^{2}}{2}+\gamma _{\sigma }\sin ^{2}\theta \right) \phi _{0}
\nonumber
\\
&+&\left( 
\frac{\gamma_{s}{}^{2}}{\rho \left( c_{s}^{2}-\upsilon ^{2}\right) }
+ \frac{
\gamma _{v}{}^{2}}{\rho \omega _{0}^{2}}\right) \phi _{0}{}^{3}=0\text{ ,}
\label{Nonlinear shordinger equation}
\end{eqnarray}
where $\rho $\ is the mass density $M/a$ , $\sin ^{2}\theta $\ represents
the positions of the protons in the hydrogen bonds and is expressed by 
\begin{equation}
\sin ^{2}\theta =\frac{1}{2}\left( 1-\frac{\bar{\varepsilon}}{\sqrt{\bar{%
\varepsilon}^{2}+t^{2}}}\right) \text{,}  \label{Sine theta square}
\end{equation}
where $\bar{\varepsilon}=\varepsilon -\gamma _{\sigma }\phi _{0}{}^{2}$ .
The interaction energy between the carriers and protons, $\gamma _{\sigma
}\sin ^{2}\theta $ , is significant only when the charge trapped by the
hydrogen bond causes proton delocalization. Therefore, for small value
of $t$ compared to $\varepsilon $\ , we can use the approximation: 
\begin{equation}
\sin ^{2}\theta \simeq 
\begin{array}{c}
0 \\ 
\frac{\gamma _{\sigma }}{2t}\phi _{0}{}^{2}-\frac{1}{2t}\left( \varepsilon
-t\right) \\ 
1
\end{array}
\begin{array}{c}
\text{ ,\ }\gamma _{\sigma }\phi _{0}{}^{2}\leq \varepsilon -t \\ 
\text{ ,\ }\varepsilon -t<\gamma _{\sigma }\phi _{0}{}^{2}\leq \varepsilon +t
\\ 
\text{ ,\ }\gamma _{\sigma }\phi _{0}{}^{2}>\varepsilon +t
\end{array}
\text{ .}  \label{Approximation of sine theta square}
\end{equation}
From the continuity of the polaron wavefunction, equation (\ref{Nonlinear
shordinger equation}) can be analytically solved and the solution is found
as combinations of Jacobian elliptic functions.

When the charge coupling with the protons is small, i.e., $\gamma _{\sigma
}\left\langle \phi _{0}\right\rangle _{\max }^{2}\leq \varepsilon -t$\
(where $\left\langle \phi _{0}\right\rangle _{\max }$\ is the maximum value
of $\phi _{0}$), the polaron is purely formed by the interaction with
phonons and the proton motion has no effect on the wavefunction or the
binding energy. The ground state solution is simply expressed as 
\begin{equation}
\phi _{0}\left( \lambda \right) =\sqrt{g}\text{ sech}\left( 2g\lambda
\right) ,  \label{Solution of hyperbolic secant}
\end{equation}
and the binding energy is 
\begin{equation}
E_{0}=-\frac{2\hbar ^{2}g^{2}}{m},
\end{equation}
where\ 
\begin{equation}
g=\frac{m}{4\hbar ^{2}}\left[ \frac{\gamma _{s}{}^{2}}{\rho \left(
c_{s}^{2}-\upsilon ^{2}\right) }+\frac{\gamma _{v}{}^{2}}{\rho \omega
_{0}^{2}}\right] .
\end{equation}
It is easy to see that $1/g$ represents the size of the polaron. To
calculate the binding energy and the size of a static polaron, we consider
that a hole carrier is doped into a poly(G)-poly(C) DNA. The effective mass
of the carrier $m=\hbar ^{2}/2t_{0}a^{2}$ is of order $20m_{e}=1.8219\times
10^{-29}$kg ($m_{e}$ the mass of a electron)\ obtained from band
structure calculations, which give a bandwidth of about $0.04$ eV \cite
{Pablo}. For the charge coupling strength $\frac{\gamma _{s}{}^{2}}{\rho
c_{s}^{2}}+\frac{\gamma _{v}{}^{2}}{\rho \omega _{0}^{2}}$ generated by the
acoustical and optical phonons in (\ref{Nonlinear shordinger equation}),
since the energy difference between the adiabatic and vertical ionization
potentials of a G-C base pair is about $0.5$ eV \cite{Hutter}, the
optical part is given as about $0.5$ eV $\times 3.38\text{\AA }=1.7$ eV \AA.
The acoustical part is assumed to be smaller, so we consider the value
of the coupling strength as approximately $2$ eV \AA . From these results,
we find that the polaron is $0.75$\AA\ wide\ with binding energy
$E_{0}=-1.35$ eV 
and, therefore, it is entirely localized in a base pair. For the
weak coupling to hydrogen bonds ($\gamma _{\sigma }\leq 0.2$ eV), the
protons are localized in their original positions and the tautomeric base
pair is unlikely to be generated.

As the coupling constant $\gamma _{\sigma }$ becomes larger, proton
transfer can be induced since the charge trapped into a base pair reduces
the bias energy $\varepsilon $\ dramatically. By considering the bias energy $%
\varepsilon $\ and the tunneling matrix element $t$ of order $0.3$ eV
\AA $^{-1}$ and $0.03$ eV \AA $^{-1}$, respectively, we find that the
binding energy of the polaron $E_{0}$ is $-1.45$ eV for $\gamma
_{\sigma }=0.226$ eV, which is more stable than the above with the
same values of the effective mass $m$ and the coupling strength $\frac{%
\gamma _{s}{}^{2}}{\rho c_{s}^{2}}+\frac{\gamma _{v}{}^{2}}{\rho \omega
_{0}^{2}}$ . The ground state wavefunction consists of two parts, the
central and the external. For the external part, the wavefunction behaves as
a hyperbolic secant function since it should vanish as $\lambda \rightarrow
\infty $ . In the central part of the polaron, the wavefunction is described
by\ a Jacobian elliptic functions of second kind\ and the proton in the
hydrogen bond is delocalized since the value of $\sin ^{2}\theta $\ is
large. For the base pair in the center of the polaron, its tautomeric form
can be easily induced.

\begin{figure} 
\hspace{-.2cm} 
\epsfxsize=8cm 
\vspace{.5cm} 
\epsfbox{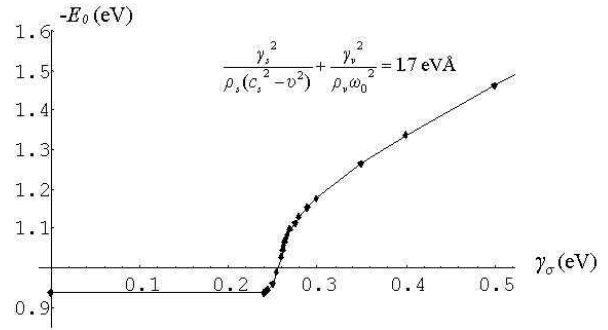} 
\caption{Polaron binding energy as a function of the coupling constant
between charges and protons $\protect\gamma _{\protect\sigma }$, where the
parameters are given by: $\frac{m}{\hbar ^{2}}=2.624$ eV$^{-1}$\AA $^{-2}$, 
$\protect\varepsilon =0.3$eV \AA$^{-1}$,\ $t=0.03$ eV \AA $^{-1}$ 
and the charge coupling strength with phonons is $1.7$ eV \AA .}
\label{E0proton} 
\end{figure}

\begin{figure} 
\hspace{-.2cm} 
\epsfxsize=8cm 
\vspace{.5cm} 
\epsfbox{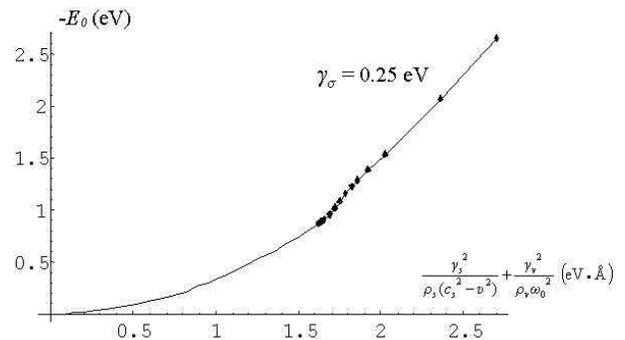} 
\caption{Polaron binding energy as a function of the coupling strength
between charges and phonons $\frac{\protect\gamma _{s}{}^{2}}{\protect\rho %
_{s}\left( c_{s}^{2}-\protect\upsilon ^{2}\right) }+\frac{\protect\gamma %
_{v}{}^{2}}{\protect\rho _{v}\protect\omega _{0}^{2}}$ , where we use the
same parameters as in Fig. 3 and the charge coupling constant with proton $%
\protect\gamma _{\protect\sigma }$ is $0.25$ eV.}
\label{E0phonon} 
\end{figure}

The relation between the coupling constant $\gamma _{\sigma }$ and the
binding energy is shown in Fig. $\ref{E0proton}$, where we have $1.7$ eV
\AA\ for the value of the coupling strength $\frac{\gamma _{s}{}^{2}}{\rho
c_{s}^{2}}+\frac{\gamma _{v}{}^{2}}{\rho \omega _{0}^{2}}$ . When $\gamma
_{\sigma }$ is larger than $0.24$ eV , the proton becomes delocalized
in the hydrogen bond and the charge binding energy increases rapidly. The
local deformations from the charge interaction with phonons can also
stabilize the polaron as shown in Fig. $\ref{E0phonon}$ with $\gamma
_{\sigma }=0.25$ eV. The increase of the couplings $\gamma _{s}$\ or $%
\gamma _{v}$\ also induces proton transfer since the carrier is strongly
attracted in the base pair.

The effect of proton transfer has been observed in experiments on the
influence of mismatches in the DNA long-range charge transfer \cite{Giese}.
Mismatched base pairs are found to cause a large drop in the charge transfer
rate due to the weakening of hydrogen bonds. For a normal G-C base pair, the
positive charge in the radical cation can be distributed by both nucleotides
through the delocalization of protons. However, mismatched Guanine, which
leads to a fast deprotonation to the surrounding water, might induce a bound
state with lower energy. Therefore, the charge is localized in the
mismatched base pair and the long-range charge transfer is interrupted.

\section{Polaron Diffusion in DNA}

The dynamic properties of a polaron induced by the carrier interaction with
acoustical and optical phonons in one dimension were explored many years ago 
\cite{Castro Neto}. In the strong coupling limit, it is found that the
polaron moves as a Brownian particle interacting with the light particles of
the environmental heat bath. To investigate these results in the context of
DNA, we first consider the case without proton coupling (i.e., $\gamma
_{\sigma }=0$).

The damping parameter of the optical polaron motion at high temperature ($%
k_{B}T\gg \hbar \omega _{0}$) is given by \cite{Castro Neto}: 
\begin{equation}
\gamma _{o}=\frac{\pi }{6}\frac{k_{B}T}{\omega _{0}}\frac{g_{o}{}^{2}}{M_{o}}%
,  \label{Gama o}
\end{equation}
where $M_{o}$\ is the classical polaron mass: 
\begin{equation}
M_{o}=\frac{m}{8}\left( \frac{E_{o}}{\hbar \omega _{0}}\right) ^{2}
\end{equation}
and 
\begin{equation}
g_{o}=\frac{m}{\hbar ^{2}}\frac{\gamma _{v}{}^{2}}{\rho \omega _{0}{}^{2}}=%
\frac{\sqrt{8mE_{o}}}{\hbar ^{2}}.
\end{equation}
Here, we use $E_{o}$ as the binding energy from the optical phonons. For
acoustical phonons we have the damping parameter: 
\begin{equation}
\gamma _{a}=\frac{315}{32\pi ^{4}}\frac{k_{B}T}{c_{s}}\frac{g_{a}}{M_{0}},
\label{Gama a}
\end{equation}
where the classical polaron mass\ for the acoustical case is 
\begin{equation}
M_{a}=\frac{32m}{3}\left( \frac{E_{a}}{\hbar c_{s}g_{a}}\right) ^{2}
\end{equation}
and 
\begin{equation}
g_{a}=\frac{m}{\hbar ^{2}}\frac{\gamma _{s}{}^{2}}{\rho _{s}c_{s}^{2}}=\frac{%
\sqrt{8mE_{a}}}{\hbar ^{2}}
\end{equation}
with $E_{a}$ the binding energy from the acoustical phonons. The random
walk motion of the polaron in the long time limit ($t\gg 1/\gamma _{o}$ or $%
t\gg 1/\gamma _{a}$) is governed by the one-dimensional diffusion equation: 
\begin{equation}
\frac{\partial n}{\partial t}=D\frac{\partial ^{2}n}{\partial x^{2}},
\label{diffusion equaiton}
\end{equation}
where $n\left( x,t\right) $ is the density of the particles and $D$ is the
diffusion coefficient. The relation between the damping parameter and the
diffusion coefficient is given by the Einstein relation: 
\begin{equation}
D_{o\left( a\right) }=\frac{k_{B}T}{\gamma _{o\left( a\right) }M_{o\left(
a\right) }},  \label{diffusion coefficients}
\end{equation}
where we use $D_{o\left( a\right) }$ for the diffusion coefficient in the
optical (acoustical) case.

As we can see from the scheme of the multiple-step hopping mechanism in Fig. 
$\ref{pdfig1}$, the hole hopping from a G-C base pair to its nearest
neighbor can be viewed as an oxidation-reduction reaction with a rate constant 
$k$ . According to the rate law of a first-order reaction, the concentration
of the charge at the $i$-th G-C base pair, $n_{i}$, is described by 
\begin{equation}
\frac{dn_{i}}{dt}=k\left( n_{i+1}+n_{i-1}-2n_{i}\right) .
\end{equation}
In the continuum limit, when the distance between two nearest neighboring
G-C base pairs, $a$, is very small, we have 
\begin{equation}
\frac{dn_{i}}{dt}=ka^{2}\left[ \frac{\frac{\left( n_{i+1}-n_{i}\right) }{a}-%
\frac{\left( n_{i}-n_{i-1}\right) }{a}}{a}\right] \rightarrow ka^{2}\frac{%
\partial ^{2}n_{i}}{\partial x^{2}}.
\end{equation}
By comparing this equation with (\ref{diffusion equaiton}), we have the
correspondence from the diffusion coefficient in the continuous medium to
the charge transfer rate of the discrete model: 
\begin{equation}
D\rightarrow ka^{2}.  \label{correspondence}
\end{equation}
Thus, instead of a diffusion coefficient, we can use the discrete
characteristic $t_{0}$ to estimate the rate constant. From (\ref{diffusion
coefficients}) and (\ref{Gama o}), we have 
\begin{equation}
D_{o}=\frac{k_{B}T}{\gamma _{o}M_{o}}=\frac{6}{\pi }\frac{\omega _{0}}{%
g_{o}{}^{2}}=\frac{3}{2\pi }\frac{\omega _{0}}{E_{o}}\frac{\hbar ^{2}}{%
2ma^{2}}a^{2}\rightarrow \frac{3}{2\pi }\frac{\omega _{0}}{E_{o}}t_{0}a^{2}.
\label{diffusion coefficients in optical case}
\end{equation}
By the correspondence (\ref{correspondence}), the rate constant for the
optical case is obtained as 
\begin{equation}
k_{o}=\frac{3}{2\pi }\frac{t_{0}}{E_{o}}\omega _{0}.
\label{rate constant for optical case}
\end{equation}
We can also derive the rate constant for the thermal activation by the
acoustical phonons as 
\begin{equation}
k_{a}=\frac{64\pi ^{4}}{315}\frac{\rho _{s}c_{s}{}^{3}t_{0}}{\gamma
_{s}{}^{2}}.  \label{rate constant for acoustical case}
\end{equation}

For poly(G)-poly(C) DNA, the oscillation frequency is $\omega _{0}\approx
10^{11}$ Hz from the theoretical and experimental results \cite
{Powell}. The theoretical calculation has shown the valence bandwidth of $%
0.04$ eV \cite{Pablo}. By considering that the binding energy $E_{o}$ is
about $1$ eV, we have the reaction rate $k_{o}\approx 10^{9} s
^{-1}$. This estimate is in agreement with the experimental result for
electron transfer in DNA which gives $k_{o} \approx 10^{8}\sim 10^{9} s
^{-1}$ \cite{Lewis}.

This low reaction rate is generated by two factors in equation (\ref
{rate constant for optical case}), the hopping probability $\frac{t_{0}}{%
E_{o}}$ and the oscillation frequency $\omega _{0}$ . Due to the strong
structural fluctuations at room temperature, the charge hopping could only
happen when the local structure of the polaron achieves an effective
configuration \cite{G Schuster}. For each period of structural oscillation,
the probability to have a successful hopping is of order $\frac{t_{0}}{E_{o}}
$ . Thus, the charge transfer rate is determined by the frequency of
reaching the effective configuration $\omega _{0}$\ and the probability $%
\frac{t_{0}}{E_{o}}$. The behavior of the incoherent charge hopping is now
described by the polaron motion under thermal fluctuations.

Another way to study the diffusion coefficent in the optical case is to
consider the mean square displacement: 
\begin{equation}
\left\langle \left( \Delta x\right) ^{2}\right\rangle =\left\langle
x^{2}\right\rangle =\int_{-\infty }^{\infty }x^{2}\phi _{0}\left( x\right)
\phi _{0}\left( x\right) dx=\frac{\pi ^{2}}{3g_{o}{}^{2}},
\end{equation}
and we may rewrite (\ref{diffusion coefficients in optical case}) as:
\begin{equation}
D_{o}=\frac{72}{\pi ^{2}}\frac{\omega _{0}}{2\pi }\frac{\left\langle \left(
\Delta x\right) ^{2}\right\rangle }{2}\sim \frac{\omega _{0}}{2\pi }\frac{%
\left\langle \left( \Delta x\right) ^{2}\right\rangle }{2}.
\end{equation}
Now, we can see that, for every oscillation period $2\pi /\omega _{0}$ , the
collisions by the light particles from the environment spread the polaron
with the mean-square displacement $\left\langle \left( \Delta x\right)
^{2}\right\rangle $. The diffusion in the continuous medium, however, needs
to be corrected when we make the correspondence since the size of the
polaron is smaller than the lattice spacing. Therefore, the real reaction
rate of charge transfer between the discrete base pairs should be smaller than
our results. Nevertheless, the order of the charge transfer rate can still
be estimated from the simple expression: 
\begin{equation}
k\sim \frac{\omega _{0}}{2\pi }\frac{\left\langle \left( \Delta x\right)
^{2}\right\rangle }{2a^{2}}.
\end{equation}

We can also estimate the reaction rate $k_{a}$ in the acoustical case from
the experimental data for the sound velocity $c_{s}\approx 10^{5}$ 
cm s \cite{Powell}. Assuming $\frac{\gamma _{s}{}^{2}}{\rho
_{s}c_{s}^{2}}\approx .34$ eV \AA , we have $k_{a}\approx 10^{13}$ s$^{-1}$ 
which is four order of magnitude larger than $k_{o}$. A faster
charge transfer rate is expected if the polaron is purely generated by the
acoustical phonons. However, if optical phonons or proton transfer are
involved, a smaller value of the reaction rate should be expected due to the
increase of the binding energy in those cases.

\section{Conclusions}

In this paper we studied the charge conduction in periodic DNA focusing
on the 
effect of the structural fluctuations which includes the proton motion in
hydrogen bonds. We found that the charge trapped by hydrogen bonds or
phonons can generate a polaron state and the coupling to the protons can
induce proton transfer in a base pair. The polaron moving with small
velocity is seen as a Brownian particle colliding with the light particles
of the environment. This diffusion process corresponds to a multiple-step
hopping mechanism in the discrete model. From this correspondence, the
reaction rate of the long-range charge transfer in DNA can be derived from
the diffusion coefficient, and it was found that the result 
predicted in the optical
case is in agreement with the experimental results.

Although we did not include the effect of proton transfer on the polaron
diffusion, it is expected to be crucial in the dynamics of charge hopping.
From electrochemical experiments \cite{Barton}, a decrease
of charge transport along the DNA with mismatched base pairs
is found. This behavior
might be explained by the effect of the proton motions in the hydrogen bonds
and we will explore this possibility in future work.

 
\section{Acknowledgements} 
 
The authors would like to thank Ward Beyermann, Douglas MacLaughlin, 
and Mike Pollak for many discussions and the  partial support provided by a
Collaborative University of California - Los Alamos (CULAR) research
grant under the auspices of the US Department of Energy. Work at Los Alamos
is supported by the US Department of Energy.


\end{multicols}

\end{document}